\documentstyle[espcrc2]{article}
\def\slashchar#1{\setbox0=\hbox{$#1$}           
   \dimen0=\wd0                                 
   \setbox1=\hbox{/} \dimen1=\wd1               
   \ifdim\dimen0>\dimen1                        
      \rlap{\hbox to \dimen0{\hfil/\hfil}}      
      #1                                        
   \else                                        
      \rlap{\hbox to \dimen1{\hfil$#1$\hfil}}   
      /                                         
   \fi}                                         %

\newcommand{ \centeron }[2]{{\setbox0=\hbox{#1}\setbox1=\hbox{#2}\ifdim
                             \wd1>\wd0\kern.5\wd1\kern-.5\wd0\fi \copy0
                             \kern-.5\wd0\kern-.5\wd1\copy1\ifdim\wd0>\wd1
                             \kern.5\wd0\kern-.5\wd1\fi}}
\newcommand{ \ltap }{\>\centeron{\raise.35ex\hbox{$<$}}
                     {\lower.65ex\hbox{$\sim$}}\>}
\newcommand{ \gtap }{\>\centeron{\raise.35ex\hbox{$>$}}
                     {\lower.65ex\hbox{$\sim$}}\>}
\newcommand{ \gsim }{\mathrel{\gtap}}
\newcommand{ \lsim }{\mathrel{\ltap}}

\def\wt{\widetilde}

\def\slashchar#1{\setbox0=\hbox{$#1$}           
   \dimen0=\wd0                                 
   \setbox1=\hbox{/} \dimen1=\wd1               
   \ifdim\dimen0>\dimen1                        
      \rlap{\hbox to \dimen0{\hfil/\hfil}}      
      #1                                        
   \else                                        
      \rlap{\hbox to \dimen1{\hfil$#1$\hfil}}   
      /                                         
   \fi}                                         %

\begin{document}
\title{Sphenomenology --- An Overview, with a Focus on a Higgsino LSP
World, and on Eventual Tests of String Theory\thanks{Invited talk at
SUSY97, Philadelphia, May 1997.}}
\author{G.L. Kane\\Randall Lab of Physics\\University of
Michigan\\Ann Arbor, MI  48109-1120}
\begin{abstract}
In this talk, as requested, I begin with a overview and with some basic
reminders about how evidence for supersymmetry in nature might appear --
in particular, how SUSY signatures are never clear so it is difficult to
search for them without major theoretical input.  Models can be
usefully categorized phenomenologically by naming their LSP -- that is,
once the LSP is approximately fixed so is the behavior of the
observables, and the resulting behavior is generally very different for
different LSPs.  Next I compare the three main LSP-models (gravitino,
bino, higgsino).  Hints from data suggest taking the higgsino-LSP world
very seriously, so I focus on it, and describe its successful prediction
of reported events from the 1996 LEP runs.  SUSY signatures in the
$\tilde h$ LSP world are very different from those that are usually
studied.  Then I briefly discuss how to measure the parameters of the
effective Lagrangian from collider and decay data.  Finally I turn to
how data will test and help extract the implications of string theories.
\end{abstract}
\maketitle

\section{INTRODUCTION}

The traditional arguments for supersymmetry continue to be compelling.
If nature is supersymmetric on the electroweak (EW) scale it provides a
solution of the hierarchy problem, it allows unification of the Standard
Model (SM) forces, local supersymmetry is connected to gravity, it
provides a derivation of the Higgs mechanism (and in that context
predicted that $M_t$ would be large), and it provides a candidate for
cold dark matter.

It is thus natural that much work has focussed on asking whether nature is
indeed supersymmetric at the EW scale.  Explicit experimental proof is
required.  Once we have that proof we have to do better, to measure the
soft-breaking terms (and eventually their phases and flavor properties)
and $\tan\beta$ and $\mu.$  The values of these parameters will point
toward the correct vacuum, and toward how SUSY is broken.

It would be very nice if clean, unambiguous experimental signals could
appear one day.  But a little thought tells us that is unlikely ---
probably impossible.  Consider colliders.  At least until the LHC, which
is unlikely to produce its first paper relevant to supersymmetry until
well over a decade from now, what will happen is that as energy and/or
luminosity increases at LEP and FNAL a few events of superpartner
production will occur. Perhaps such events have already occurred.  Each
event has two escaping LSP's , so it is never possible to find a
dramatic $Z$-like two body peak, or even a $W$-like peak with one
escaping particle. Even worse, often several channels look alike to
detectors so simple features can be obscured.  And usually there are SM
processes that can fake any particular signature, as well as ways to
fake signatures because detectors are imperfect.

Thus to make progress it is essential to proceed with limited amounts of
incomplete information.  Without theory input it is entirely possible
that signals would not be noticed, hidden under backgrounds since there
was no guide to what cuts to use (an example is discussed below).
Further, a particular signal might be encouraging but not convincing --
only when combined with other signals that were related by the theory
but not directly experimentally could a strong case be made.

What about information from decays?  In the best cases, such as $b\to
s\gamma$, where there is no tree level decay, the SUSY contribution
could be comparable to the SM one, say a large effect of order 30\%.
Then to get a significant effect the combined errors of the theoretical
calculation of the SM value and the experiment have to be below 10\%.
Only $b\to s\gamma$ of known decays can approach that; presently the
theoretical error\cite{misiak} is at about that level, and the
experimental error about 20\%.  At best, in a couple of years this could
provide compelling evidence of new physics.  If so, taken alone several
interpretations would be possible, but with theoretical input it could
be combined with collider data to determine which were consistent.

When there is a tree level SM contribution, such as for $R_b =
\Gamma\left(Z\to b\bar b\right)/ $ $\Gamma\left(Z\to\>
\hbox{hadrons}\right)$ the SUSY effect has to be a loop and can be $\sim
{1\over 2}$\%, so experimental errors have to be several times smaller.
Just from statistics that requires $\sim 10^6$ events, which is
unlikely.  Sometimes several decays are related in a particular model,
in which case the combined predictions can be tested and the results are
somewhat more significant (that is the situation for $b\to s\gamma$ plus
$R_b$ plus $\alpha_s)$.  Further input could come from proton decay to
channels favored by SUSY, from occurrence of decays forbidden in the SM
such as $\mu\to e\gamma$ or $K\to \mu e$, from neutron or electron
electric dipole moments, from non-SM CP violation, or other rare
phenomena.

If we can discover superpartners before LHC, it will be necessary to
proceed with fragments of information, check their consistency, make
predictions to test them, and slowly build a case.  We will see below
that there are things to work with (which need not have happened -- most
fluctuations from the SM could never be interpreted as SUSY).  There is
now some evidence for one prediction based on those hints.  A number of
predictions at LEP and FNAL can test whether this interpretation, based
on the higgsino-LSP world, actually describes nature.

\section{COMPARISONS}

Almost all of the studies of supersymmetric models can be classified
according to what is the LSP, and fall in three categories, as show in
Table 1.  On the left are listed several criteria that are often used to
compare and test models.  The first world listed has a light gravitino
LSP ($\widetilde{\rm G}$LSP), the second an LSP that is mainly bino
($\widetilde{\rm B}$LSP) and the third mainly higgsino ($\widetilde
h$LSP).  The implications for SUSY-breaking and experimental signature
are very different for the three cases -- it will be easy to recognize
which is being observed once one is detected.  ($\widetilde{\rm G}$LSP)
corresponds to gauge-mediated SUSY-breaking, and the other two
gravity-mediated SUSY-breaking.)  Some of the individual criteria will be
described in more detail below in the ($\widetilde h$LSP)
section; see also ref. 2.

\def\glsp{\widetilde{\rm G}LSP}
\def\blsp{\widetilde{\rm B}LSP}
\def\hlsp{\widetilde{\rm h}LSP}

\begin{table*}
\begin{tabular}{l c c c}
&\multicolumn{3}{c}{\parbox[t]{4in}{\ \ \ \ 
\ \ \ \ \ \ \ \ \ \ \ \ Table 1.}}\\[8pt]
Evidence/Criteria&$\glsp$&$\blsp$&$\hlsp$\\[8pt] \hline
&&& \\[-1pt]
Absence of FCNC&Yes, if messenger scale low&\multicolumn{2}{l}
{\parbox[t]{2.5in}{Mechanisms exist, but don't know if they are 
applicable}} \\ 
&&& \\[-1pt]
Number of parameters&\multicolumn{3}{l}{\ \ \parbox[t]{4.2in}{Presently all 
about same -- $\glsp$ somewhat more than others now, 
but will be more predictive after squarks and sleptons observed.}} \\
&&& \\[-1pt]
{\parbox[b]{1.5in}{CDF $e$ \lq\lq $e$''
$\gamma\gamma\slashchar{E}_T$}}&maybe&no&yes \\
&&& \\[-1pt]
$R_b, BR(b\to s\gamma), \alpha_s$&\parbox{.8in}{maybe
bs$\gamma$\hfil\break 
no for $R_b, \alpha_s$}&no&yes \\
&&& \\[-1pt]
LEP\ $\gamma\gamma\slashchar{E}$ events&no&no&yes \\
&&& \\[-1pt]
Cold Dark Matter&not LSP&ok&yes \\
&&& \\[-1pt]
$\tilde g,$ light $\tilde t$ at FNAL?&no&no&ok \\
&&& \\[-1pt]
EW baryogenesis&no&no&ok \\
&&& \\[-1pt]
$m_{h^\circ} \lsim M_Z$&no reason&no reason&yes \\
&&& \\[-1pt]
{\parbox{1.2in}{typical signatures\hfill\break
that distinguish}}&\parbox[t]{1.3in}{(a) either 2 $\gamma$'s in every
event; or 2 charged leptons in every event; or long-lived particles that
decay in or outside of detector\hfill\break
(b) medium $\slashchar{M}$}&\parbox[t]{1.5in}{(a) no $\gamma$'s;\hfill\break (b)
events with no charged leptons;\hfill\break 
(c) larger $\slashchar{M}$\hfill\break
(d) trilepton + $\slashchar{E}$ events at FNAL, LHC}&\parbox[t]{1.6in}{(a)
0, 1, or 2 $\gamma$'s\hfill\break (b) large $\slashchar{M}$
(c) not only LSP but also $\tilde \nu , \tilde N_3$ invisible} \\  
\end{tabular}
\end{table*}

In a $\glsp$ world there is a small window for a signal at LEP but no
reason to expect one there.  If the CDF event\cite{park} were
interpreted as evidence for a $\glsp$ world, which is difficult given
constraints but not excluded, many such events would occur at FNAL after
the collider starts running again in 1999; otherwise there is only a
small window at FNAL.  Such a world could probably be detected at LHC or
a lepton collider about 2010.

In a $\blsp$ world there is no evidence today for sparticles, and all
hints must disappear (no more $ee\gamma\gamma\slashchar{E}_T$ events at
FNAL; $BR (b\to s\gamma) \to$ SM, $R_b \to SM,$ $\alpha^{\Gamma_Z}_s -
\alpha^{\rm other}_s \to 0;$ no future excess of $\gamma\gamma
\slashchar{E}$ events at LEP; baryogenesis not at EW scale; etc.).
There are small windows at LEP and FNAL but no reason for sparticles to
be there.  Such a world probably be detected at LHC or a lepton collider
about 2010.

In a $\hlsp$ world sparticles may have already been observed.
Confirmation will occur at LEP\cite{mahlon} once 50
pb$^{-1}$/detector at $\sqrt{s} \gsim 190$ GeV has been accumulated or
before.  In the next section the hints for an $\hlsp$ world and some of
the tests are described in a little more detail.

\section{$\hlsp$ WORLD}

Here we will need some notation.  $\widetilde N_i$ are the four
neutralino mass eigenstates, $\widetilde C_i$ the two chargino ones, and
$\tilde t_1$ the lighter stop mass eigenstate.  $M_1$ and $M_2$ are the
$U(1)$ and $SU(2)$ soft-breaking gaugino masses, $\mu$ the coefficient
of $H_U H_D$ in the superpotential, and $\tan\beta$ the ratio
$<H_U>/ < H_D>.$

All the phenomenological analysis can be done with a general
soft-breaking effective Lagrangian written at the electroweak scale.
One can fully analyze the CDF event, the LEP $\gamma\gamma
\slashchar{E}$ events, $b\to s\gamma,$ $R_b, \alpha_s,$ EW baryogenesis
and cold dark matter with only 5 major parameters ($\mu , \tan\beta,
M_1, M_2, M_{\tilde t_1}$).  The complete detailed analysis also depends
on the stop section mixing angle, and the sneutrino and $\tilde e_R$
masses but not sensitively.  The other parameters of the Lagrangian will
enter once more sparticles are being detected.

I don't have space here to give details about most of the entries in the
table; a recent summary is available in ref. 2.  Here I will only mention
the LEP $\gamma\gamma\slashchar{E}$ events since they have had less
exposure.  In 1986 we argued\cite{haber} that collider events with hard
isolated photons would be a good signature for supersymmetry, coming
from $\wt \gamma \to \gamma \wt h.$ In Jan. 1996 we learned about the
CDF event, and analysis\cite{sandro} showed it could indeed be
interpreted as a SUSY candidate.  Further analysis showed
consistency\cite{wells,wellstwo,mrenna} (same parameters) with other
hints of evidence for SUSY.  Consequently, we predicted\cite{sandro}
that several confirming channels could show up at LEP or FNAL.

One of the confirming channels\cite{mahlon} is events with two hard
$\gamma$'s at LEP, from $e^+e^- \to \widetilde N_2\widetilde N_2$
followed by $\widetilde N_2 \to \gamma \widetilde N_1.$ $\wt N_2$ is
mostly $\wt\gamma$ and $\wt N_1$ mostly $\wt h.$ Such events must
have missing invariant mass $\slashchar{M}$ above $2M_{LSP} \gsim M_Z$;
since there is a large background with $\slashchar{M} \simeq M_Z$ one
should make a cut with $\slashchar{M} \gsim$ 100 GeV, which loses little
or no signal.  Given the motivation from other data there should be a
minimum $E_\gamma \gsim 5$ GeV so a cut at (say) 5 GeV for both photons
gets rid of most of the soft $\gamma$ background.  The signal is
isotropic while the background peaks sharply near the beam so a cut of
(say) $|\cos\theta | < 0.85$ gets rid of a large background at a
relatively small loss in signal.

About 6 candidates have been reported for 161 + 172 GeV, 4
detectors.\cite{wilson} The \lq\lq about'' is needed since the full
parameters of the events have not been reported.  With these cuts the
background cross section\cite{ambros} is about 0.025 pb, so at 161 + 172
GeV for 4 detectors the background is about 1.5 events for detection
efficiency of 3/4.  This number assumes an increase of about 35\% over
the tree level for radiation of extra undetected $\gamma$'s.  No general
background calculations have been completed so these numbers are
estimates based on the information in ref. 11.

While not yet compelling, such a result is certainly encouraging.
Mahlon and I have catalogued\cite{mahlon} several tests that will be
carried out at LEP as soon as it gets near its design luminosity at
higher energies.

In the absence of one very convincing channel, we have to look at the
pattern of several processes as I explained in the introduction.  This
is in a good tradition -- in the 1970's checking that $\sin^2\theta_W$
was consistent with a single value when measured different ways was one
of the main ways we gained confidence in the EW theory.  Much earlier,
checking that Avagadro's number had the same value when measured many
different ways was considered the strongest argument for the physical
existence of atoms near the beginning of this century.\cite{perrin}

\def\wt{\widetilde}
Each of the experimental hints I mentioned is about as strong as it
could be given existing integrated luminosities and experimental
errors.  Each can only be interpreted as evidence for SUSY if the
parameters $(\mu , \tan\beta, M_1, M_2, M_{\tilde t_1})$ take on a small
range of values.  If we were being misled by the data, if these
phenomena were not evidence for SUSY, it is very unlikely that all would
have given consistent quantitative descriptions for the \underbar{same}
parameters.  Very encouraging. Conservatively, finding the same
parameters is a necessary condition, but of course not sufficient.

{Signatures at LEP and FNAL\cite{mahlon} are interesting
and rather different from the usual ones.  $\tilde \nu$ is light, and
dominantly decays $\tilde\nu \to \nu\widetilde N_1$ so it is mainly
invisible.  $\wt N_2 \to \gamma \wt N_1$ as discussed above, giving
$\gamma\gamma\slashchar{E}$ events from $e^+e^- \to \wt N_2 \wt N_2.$ $\wt
N_3 \to \nu\tilde\nu$ dominates so $\wt N_3$ is mainly invisible.  $\wt
C^\pm_1 \to \ell^\pm\tilde\nu$ dominates, and $m_{\wt C_1} -
m_{\tilde\nu}$ is small so the leptons can be very soft.  Production of
$\wt N_2 \wt N_3$ gives events with one $\gamma$ and large missing
energy, and several other channels also contribute to this signature.
All of these channels can be seen at LEP but could be missed if
appropriate cuts and analyses were not made.  In particular, for
$\gamma\gamma \slashchar{E}$ events if photons softer than $\sim$ 5 GeV
are included the background will increase rapidly and a signal could be
hidden.}

\section{FROM DATA TO ${\cal L}_{EFF}$}

Actual measurements of effects of superpartners will produce cross
sections and distributions, excesses of events with some set of
particles such as gammas and perhaps with missing energy.  They do not
produce measurements of the masses or couplings of superpartners, and
determination of the soft-breaking parameters or $\mu$ or $\tan\beta$ is
even less likely.  Most distributions get contributions from several
processes as well.  How can we proceed to extract the physics parameters
of interest from data in cush a nonlinear situation?

Some analyses already exist \cite{feng}, mainly for $e^+e^-$ colliders
(where results are often simpler) rather than hadron colliders, and for
a few idealized cases.  These already show that useful general results
can be obtained in such situations.

In fact, the general problem has been addressed and a procedure
given\cite{kribs} to extract the sparticle masses and/or the parameters
of ${\cal L}_{\small EFF}$.  Every measurement provides information.
There is an excess of events at a certain cross section level in one
process, none in another.  The optimum procedure is to randomly select
values for all parameters, calculate all observables, and discard values
of parameters that give observables in disagreement with data.  While
that sounds like a big task it is not so hard in practice since any
given observable only depends on a few parameters. Obviously to
determine $N$ parameters one will need about $N$ observables sensitive
to the parameters.  In practice the theory provides strong constraints
so the actual number needed can be smaller than $N$.  Also, often it is
easy to put limits on parameters that reduce the size of the problem
after a little thought.  This method could already be used to get
general limits on sparticle masses, but so far only parameter dependent
limits have been published as far as I know.  This procedure has been
used in references 4, 6.

\section{MEETING AT THE UNIFICATION SCALE}

As data about superpartners is increasingly available, more and more of
the parameters of the effective Lagrangians ${\cal L}_{EFF}$ of the
theory at the electroweak scale will be measured.  Constraints from rare
decays, CP violation, baryogenesis etc. will be included.  Then,
assuming the theory to be perturbative to the scale where the gauge
coupling unify, the effective Lagrangian at that scale will be
calculated by using renormalization group equations.\cite{carena} It is
not necessary (nor expected) that there be a desert in between, but only
that the theory be perturbative.  Intermediate matter and scales are
expected.  There will be consistency checks that allow the
perturbativity to be confirmed.  Since constraints occur at both ends it
is not just an extrapolation.  Unification may or may not involve a
unified gauge group.

String theory, on the other hand, starts somewhat above the unification
scale, at the Planck scale.  If the way to select the vacuum was known,
and also how SUSY was broken (perhaps once the vacuum is known the latter
will be determined), then the ${\cal L}_{EFF}$ could be predicted, and
compared with the ${\cal L}_{EFF}$ deduced from data.  In  practice I
expect it to be the other way, as it has been throughout the history of
physics -- once we know the experimental ${\cal L}_{EFF}$ deduced from
data the patterns of parameters will be recognizable and will tell
someone how SUSY is broken and how the vacuum is selected.  After that
it will be possible to derive it from string theory.  String theorists
and sphenomenologists will meet at the unification scale.

In the $\hlsp$ world some interesting preliminary results have been
obtained (by the method of the previous section). Whether these results
persist as better data is obtained or not, they encourage one to think
we will be able to learn about unification scale physics from EW scale
data.  Some examples:

(a)  The usual \lq\lq unification'' assumption for gaugino masses has
$M_1\!=\!M_2$ at the unification scale.  Then the RGE running gives
$M_1(M_Z) = {5\over 3} \tan^2 \theta_W M_2 (M_Z)$ $\approx {1\over 2} M_2
(M_Z).$    If the CDF event and/or the LEP $\gamma\gamma\slashchar{E}$
events are indeed production of superpartners in a higgsino LSP world,
then we know $BR (\widetilde N_2 \to \widetilde N_1 \gamma)$ is large.
Examination of the neutralino mass matrix\cite{mele} shows that this $BR$ is
maximal when $M_1 (M_Z) \cong M_2 (M_Z)$, and for it to be $\gsim {3\over
4}$ one needs $M_1 (M_Z) > {1\over 2} M_2 (M_Z)$.  This in turn implies
that the simple unification condition does not hold, which means we are
learning about unification scale physics from collider data.  A similar
result for $M_1 (M_Z)/M_2 (M_Z)$ has been found in ref. 16.

Note that it will be interesting to understand how to interpret  this
result.  It could directly point toward a theory where $M_1 (M_U) > M_2
(M_U).$  Or it could, say, imply that the true neutralino mass matrix is
of the form

$$\pmatrix{M'_1\cr &M_1\cr &&M_2\cr &&&\ddots\cr}$$
\bigskip

{\noindent because of an extra $U(1)$. Then the trace of this is $M'_1 +
M_1 + M_2 + \ldots$ so if we have an effective $4\times 4$ neutralino
mass matrix what we call $M_1$ is really $M_1 + M'_1 > M_1.$}

(b) Most of this phenomenology suggests $\tan\beta$ is near or below its
naive perturbative lower limit. But that is subtle.  It depends
sensitively on $M_{\rm top}$ (if $M_{\rm top}$ decreases from 175 GeV to
163, the naive lower limit on $\tan\beta$ decreases from 1.76 to 1.38),
and the value of $M_{\rm top}$ should not be taken as settled yet.
SUSY-QCD effects\cite{polonsky} lower the limit below the naive one.
Also, new physics at intermediate scales could affect this lower limit
which would be very interesting.

(c) In the MSSM, if $\alpha_s (M_Z) \lsim 0.125$ then sparticles must be
heavier than about 1 TeV if the three gauge couplings exactly
unify.\cite{pierce} But the world average is $\alpha_s = 0.117 \pm
0.003$ (for consistency this should be calculated leaving out the
$\alpha_s$ from $\Gamma_Z$). So if there is any evidence for sparticles
then some additional physics must affect the running of the gauge
couplings.  A number of possibilities exist and it will be very
interesting to elucidate which one(s) occur.

(d) If a new $U'_1$ symmetry exists, with a $Z'$ at the TeV scale, the
$Z'$ may be too heavy and too weakly coupled to quarks and leptons (much
of its coupling may be to sparticles and heavy Higgs bosons) to detect
directly.  But it affects squark and slepton masses\cite{kolda} through
$D'-$terms, and through them one can determine the $U'_1$ charges.  IF
all the details of the CDF event and LEP data were not misleading us
then probably $\tilde e_R$ is heavier than $\tilde e_L,$ which suggests
a non-minimal contribution to the slepton masses that could come from
$D'-$terms.

\section{TESTING STRING THEORY}

A myth has grown that string theories are not normal testable physics.
The myth began in the middle 1980's, partly as a reaction to excessive
enthusiasm from proponents of string theory.   We have learned a lot
since then.

The myth is wrong.  String theory is testable normal science.  You don't
have to go somewhere in space or time to test a theory about there.  We
know many examples. The big bang theory is well tested by its correct
predictions now of the expansion of the universe, nucleosynthesis, and
the cosmic microwave background radiation even though we can't go back
to observe it.  The composition of distant stars, the facts that they
are made of the same atoms as our star and us, and that they obey the
same rules of quantum theory and relativity as hold in our part of the
universe, can be learned by understanding spectra and redshifts and
making appropriate observations without going there.  We can learn how
the dinosaurs became extinct without watching them die.

It is always crucial to go through chains of reasoning and to do
complicated calculations in order to perform the tests.  That is true of
all the above examples, and equally true of physics today.  Condensed
matter physicists believe they have a quantitative understanding of
phase transitions.  That belief is based on complicated renormalization
group analyses, involving exactly the same techniques used to connect
physics at the electroweak scale to physics at the Planck scale -- it is
even true in both cases that at the place one wants most to study the
theory is expected to become non-perturbative.  Parity violation in
atoms is thought to be a quantitative test of the SM even though very
complicated calculations are needed to go from observing an effect to
extracting the couplings of the $Z$ to quarks and leptons.  Calculations
will also be needed in the string case but that does not mean it is less
testable than other normal physics.

String theory is not yet very tested or predictive.  The techniques to
change that are being learned.  As always in physics, one proceeds by
making assumptions and calculating, and slowly improving the results.
Normally it takes a while to get it right.

In the following I give some examples to make concrete how many tests
there will eventually be of string theory.  As is well understood, it is
necessary to have not only the theory, but also to know the vacuum.
Presumably determining the correct vacuum also determines how
supersymmetry is broken and vice versa, but in practice there might be
progress in one or the other of these first.  As I said earlier, I expect
data will be necessary to make progress on these.  

It is  useful to give examples in six categories.  Note that no
super-high energy facilities are needed for any of the examples.  I list
a number of examples; more will exist.  In order to not give dozens of
references here I will give none, with apologies to the many people who
have discussed these ideas and observables.

(a) Profound questions.  If a theory can provide a definition of
space-time, or derive the existence of three and only three non-compact
space dimensions, or explain what a particle \underbar{is}, or explain
what electric charge \underbar{is}, or explain the value of the
cosmological constant, or solve the black hole information loss puzzle,
that theory has passed a major test.

(b)  Why questions.  Why are there three chiral families of quarks and
leptons?  Why is general relativity the correct theory describing gravity
(or whatever is)?  Why is nature supersymmetric?  Why is the SM gauge
group $SU(3) \times SU(2) \times U(1)$?  Why is matter quarks and leptons
but not leptoquarks?

Every answer to a \lq\lq profound question'' or \lq\lq why question'' is
a powerful test of a theory.  Many of these questions are true stringy
ones rather than ones likely to hold in any high scale theory.

(c)  Very low energy phenomena -- no collider needed.  String theories
provide expressions for fermion masses, though so far they cannot be
evaluated.  Eventually there will be a calculation of (say)
$m_\mu/m_\tau$ (a useful ratio because it is not too sensitive to small
corrections).  Once the quark mass matrix can be written in the
$SU(2)\times U(1)$ basis it can be diagonalized and the CKM angles,
including the weak phase, calculated.  A string theory will predict
whether the proton decays, and if so its lifetime and branching ratios;
the decay may be forbidden by symmetries even if the theory has a unified
gauge group.  Neutrino masses and mixing angles come from physics beyond
the SM and should be predicted.  The strong CP problem should be
explained by a string theory.  Forbidden decays such as $\mu \to
e\gamma$ and $K\to \mu e$ may be induced and their rates predicted.  The
phases of the soft-breaking terms may be learned from the electric
dipole moments of the neutron and electron, and from $\epsilon_k,
\epsilon_B,$ and $B_d, B_s$ mixing.  The baryon asymmetry of the
universe provides information on magnitudes and phases of soft-breaking
masses and $\mu$ and $\tan\beta$.  Possibly laboratory cold dark matter
experiments will measure the mass and couplings of the LSP.

(d)  Collider phenomena.  Once superpartners are being studied we will
have over 30 of their masses to test any theory of SUSY-breaking.  In
addition there will be a number of SCKM angles and phases, though
perhaps it will be a long time before all of them can be measured.
Branching ratios and missing energy events will tell us that $R$-parity
and perhaps other discrete symmetries are conserved (or not).  The gauge
couplings and the scale at which they unify are sensitive to any
predicted intermediate scale matter and to non-renormalizable operators
with inverse powers of $M_{Pl}.$  Extra $U'_1$'s could show up as
$Z$'s (which may be hard to detect, see above) or as $D'$-terms
affecting squark and slepton masses.

(e) Cosmology.  The scalar potential that contains the inflaton(s) and
drives inflation(s) is the supersymmetric scalar potential, containing a
number of scalar fields with and without SM quantum numbers.  It has
flat directions lifted by soft-breaking terms and Planck scale
operators.  The parameters in that potential not only determine the
course of inflation, they also affect or determine $\nu$ masses, the
baryon asymmetry, structure formation, scollider physics, gravitational
waves from before the big bang, axions, etc.  The dark matter from a
supersymmetric unified theory is expected to contain cold dark matter
determined by the LSP, and possibly cold dark matter from axions arising
from breaking global symmetries of the theory, hot dark matter from
neutrino masses, and baryons.  The amounts and ratios of these should
eventually be calculable.

(f)  Unexpected.  In the past all major theoretical progress has led to
unanticipated predictions that were major tests, and that is likely here
too.  Examples are the prediction of antiparticles from unifying special
relativity and quantum theory, and electromagnetic waves from unifying
electricity and magnetism.  Obviously I cannot say what will arise
here.  Perhaps there will be long range forces, or the vacuum will not
respect Lorentz invariance of CPT -- such phenomena have been suggested.

\section{SUMMARY}

\begin{description}
\item[$\bullet$] It may be hard to recognize signatures of supersymmetry
at colliders without theoretical input.  The standard ones used in
studies are all for $\blsp$.  The main signatures can be very different
for $\glsp$ and $\hlsp$.

\item[$\bullet$] We understand how to extract the basic soft-breaking
parameters of the Lagrangian from data containing effects of
superpartners, but theory is essential -- there are no
theory-independent signals.

\item[$\bullet$] Maybe superpartners are already being observed in
several ways.  If so, we live in a $\hlsp$ world.  If so, LEP and FNAL
are the main facilities for particle physics of the 21$^{\rm st}$
century -- higher luminosity and better detectors are the main need for
FNAL.

\item[$\bullet$] To find out, watch at LEP2 for events with large
missing invariant mass $(\gsim 100 {\rm GeV})$ with signatures
$\gamma\gamma\slashchar{E}, \gamma\slashchar{E}, \ell^\pm \ell'^\mp
\slashchar{E}\ldots$.

\item[$\bullet$] In a $\hlsp$ world we expect $h^\circ$ to be observed
at LEP, and it certainly can be observed at FNAL.

\item[$\bullet$] Electroweak scale data on superpartners will constrain
the effective Lagrangian at the unification (GUT or string) scale and
operators $\sim M_{\rm unif}/M_{Pl},$ leading to insights about SUSY
breaking and the vacuum structure.

\item[$\bullet$] String theory is testable, normal science.
\end{description}

\section{Acknowledgements}

I have benefitted from discussions with S. Ambrosanio, M. Carena,
M. Einhorn, S. Eno, H. Frisch, T. Gherghetta, A. Kostelecky, G. Kribs,
G. Mahlon, S. Martin, S. Mrenna, R. Pain, D. Stickland, S. Ting,
D. Treille, G. Wilson, and E. Witten.

\end{document}